\documentclass[english]{article}
\usepackage[LGR,T1]{fontenc}
\usepackage[latin9]{inputenc}
\usepackage{textcomp}
\usepackage{amsmath}
\usepackage{amssymb}
\usepackage{graphicx}
\usepackage{esint}
\usepackage[left=1.75cm,top=1.5cm,right=1.75cm,bottom=2.25cm]{geometry}
\begin{document}
\begin{center}
\Large{\bf{Anisotropic String Cosmological Models in Heckmann-Suchuking Space-Time}}\\
\vspace{4mm}
\normalsize{G. K. Goswami$^1$, R. N. Dewangan$^2$,
A. K. Yadav$^3$ \& A. Pradhan$^4$}\\
\vspace{2mm}
\normalsize{$^{1,2}$Department of Mathematics, Kalyan P. G. College, Bhilai - 490006, India}\\
\vspace{1mm}
\normalsize{Email: gk.goswam9@gmail.com}\\
\vspace{2mm}
\normalsize{$^3$Department of Physics, United College of Engineering $\&$ Research, Greater Noida - 201306, India}\\
\vspace{1mm}
\normalsize{Email: abanilyadav@yahoo.co.in}\\
\vspace{2mm}
\normalsize{$^{4}$Department of Mathematics, GLA University, Mathura, India}\\
\vspace{1mm}
\normalsize{Email: pradhan@iucaa.ernet.in}\\
\end{center}
\vspace{2mm}
\begin{abstract}
In the present work we have searched the existence of the late time acceleration
of the universe with string fluid as source of matter in anisotropic Heckmann-Suchking space-time by
using 287 high red shift $(0.3 \leq z\leq 1.4)$ SN Ia data of observed absolute magnitude
along with their possible error from Union 2.1 compilation. It is found that the best fit values
for $(\Omega_{m})_{0}$, $(\Omega_{\Lambda})_{0}$, $(\Omega_{\sigma})_{0}$ and $(q)_{0}$ are 0.2820, 0.7177, 0.0002 $\&$
-0.5793 respectively. Several physical aspects and geometrical properties of the model are discussed in detail.\\
\end{abstract}
\section{Introduction}
SN Ia observations (Riess et al. 1998; Perlmutter et al. 1999) confirmed that the observable
universe is undergoing an accelerated expansion. This acceleration is realized due to unknown cosmic fluid -
dark energy (DE) which have positive energy density and negative pressure.
So, it violate the strong energy condition (SEC). The authors of ref. [3] confirmed that
the violation of SEC gives anti gravitational effect that provides an elegant description
of transition of universe from deceleration zone to acceleration zone.
In the literature, cosmological constant is the simplest candidate to describe the present acceleration of universe
(Gr\O{}n and Hervik 2007) but it suffers two problems - the fine tunning and cosmic coincidence problems
(Carroll et al. 1992, Copeland et al. 2006). In the early of 21 century, some authors (Copeland et al.  2006, Alam et al. 2003)
 have developed few scalar field DE models, namely the phantom, quintessence and
k-essence rather than positive $\Lambda$. In the physical cosmology, the dynamical form of DE with an
effective equation of state (EOS), $\omega < -\frac{1}{3}$, were proposed instead of positive $\Lambda$
(Komastu et al. 2009, Riess et al. 2004, Astier et al. 2006).\\

The substantial theoretical progress in string theory has brought
forth a diverse new generation of cosmological models, some of which are subject to
direct observational tests. Firstly the gravitational effect of cosmic strings are investigated by Letelier (1979, 1983).
The present day observations of universe indicate the existence of large scale
network of strings in early universe (Kible, 1976, 1980).
Recently, Pradhan et al. (2007) and Yadav et al. (2009) have studied string cosmological model in
non-homogeneous space-time in which geometric strings were considered as source of matter.
After the big bang, the universe may have undergone so many phase transitions as its temperature
cooled below some critical
temperature as suggested by grand unified theories (Zel'dovich et al. 1975; Kibble 1976, 1980;
Everett 1981; Vilenkin 1981). At the very early stage of evolution
of universe, the
symmetry of universe was broken spontaneously that rises some topologically-stable defects such as
domain walls, strings and monopoles (Vilenkin 1981, 1985). Among all the three cosmological structures,
only strings give rise the density perturbations that leads
the formation of galaxies. Pogosian et al. (2003, 2006) have suggested that the cosmic strings are not
responsible for CMB fluctuations and observed clustering of galaxies.   In 2011, Pradhan and Amirhashchi (2011) have considered time varying scale factor that generates a transitioning universe
in Bianchi -V space-time. Yadav et al. (2011) and
Bali (2008) have obtained Bianchi-V string cosmological models in general relativity.
The magnetized string cosmological models are discussed by some authors (Chakraborty 1980; Tikekar and Patel
1992, 1994; Patel and Maharaj 1996; Singh and Singh 1999; Saha and Visineusu 2008)  in different physical contexts. Recently Goswami et al (2015) have
studied $\Lambda$CDM cosmological model in absence of string. In this paper
we have established the existence of string cosmological model in anisotropic
Hecmann-Suchuking space-time.The organization of
the paper is as follows: The model and field equations are presented in section 2. In Section 3, 4  $\&$ 5
we obtain the expressions for Hubble constant, Luminosity distance and apparent magnitude. Section 6 deals the estimation
of present values of energy parameters. Finally the conclusions of the paper are presented in section 7. \\
\section{The Model and Field equations}
We consider a general Heckmann-Schucking metric
\begin{equation}\label{1}
  ds^{2}= c^{2}dt^{2}- A^{2}dx^{2}-B^{2}dy^{2}-C^{2}dz^{2}
  \end{equation}
  where A, B and C are functions of time only.\\
  The energy-momentum tensor for a cloud of massive strings and perfect fluid distribution is taken as
 \begin{equation}\label{2}
    T_{ij}=(p+\rho)u_{i}u_{j}-pg_{ij}-\lambda x^i x^j
 \end{equation}
  with
  \begin{equation}\label{3}
    g_{ij}u^{i}u^{j}=1
  \end{equation}
  and
  \begin{equation}\label{4}
     x^ix^i = -1,u^{i}x_i=0
  \end{equation}
where $u^{i}$ is the four velocity vector, $\rho$ is the rest energy density of the system of
strings, ~$\lambda$ is the tension density of the strings and~ $x^{i}$ is a unit space-like vector
representing the direction of strings.\\
In co-moving co-ordinates,
  \begin{equation}\label{5}
    u^{\alpha}=0,\alpha=1,2 or 3.
  \end{equation}
Choosing $x^i$ parallel to $\partial/\partial x$, we have
\begin{equation}\label{6}
    x^i = (A^{-1}, 0, 0, 0).
\end{equation}
If the particle density of the configuration is denoted by $\rho_m$, then
\begin{equation}\label{7}
    \rho=\rho_m +\lambda
\end{equation}
The Einstein field equations are
  \begin{equation}\label{8}
    R_{ij}-\frac{1}{2}Rg_{ij}+ \Lambda g_{ij}= -\frac{8\pi G}{c^{4}}T_{ij}
  \end{equation}
  Choosing co-moving coordinates,the field equations (8) in terms of line element (1) can be written as
  \begin{equation}\label{9}
    \frac{B_{44}}{B}+\frac{C_{44}}{C}+\frac{B_{4}C_{4}}{BC}=-\frac{8\pi G}{c^{2}} (p-\lambda)+ \Lambda c^{2}
  \end{equation}
   \begin{equation}\label{10}
    \frac{A_{44}}{A}+\frac{C_{44}}{C}+\frac{A_{4}C_{4}}{AC}=- \frac{8\pi G}{c^{2}}p+\Lambda
    c^{2}
  \end{equation}
  \begin{equation}\label{11}
   \frac{A_{44}}{A}+\frac{B_{44}}{C}+\frac{A_{4}B_{4}}{AB}=- \frac{8\pi G}{c^{2}}p+\Lambda
   c^{2}
  \end{equation}
   \begin{equation}\label{12}
     \frac{A_{4}B_{4}}{AB}+\frac{B_{4}C_{4}}{BC}+\frac{C_{4}A_{4}}{AC}= \frac{8\pi G}{c^{2}}\rho+\Lambda
     c^{2}
   \end{equation}
   where $A_{4}$,$B_{4}$ and $C_{4}$ stand for time derivatives of A,B,C respectively. the mass-energy conservation equation $T^{ij}_{;j}=0$ gives
   \begin{equation}\label{13}
     \rho_{4}+(p+\rho)(\frac{A_{4}}{A}+\frac{B_{4}}{B}+\frac{C_{4}}{C})-\lambda \frac{A_{4}}{A} =0
   \end{equation}
   Where $A_{4}$,$B_{4}$ and $C_{4}$ stand for time derivatives of A,B and C respectively. Subtracting (10) from (9),(11) from (10) and (11) from (9) we get
   \begin{equation}\label{14}
    \frac{A_{44}}{A}-
    \frac{B_{44}}{B}+\frac{A_{4}C_{4}}{AC}-\frac{B_{4}C_{4}}{BC}=-\frac{8\pi G}{c^{2}}\lambda
   \end{equation}
   \begin{equation}\label{15}
     \frac{B_{44}}{B}-
     \frac{C_{44}}{C}+\frac{A_{4}B_{4}}{AB}-\frac{A_{4}C_{4}}{AC}=0
\end{equation}
\begin{equation}\label{16}
     \frac{C_{44}}{C}-
     \frac{A_{44}}{A}+\frac{B_{4}C_{4}}{BC}-\frac{A_{4}B_{4}}{AB}=\frac{8\pi G}{c^{2}}\lambda
\end{equation}
Subtracting(16) from (14), we get
\begin{equation*}
\frac{B_{44}}{B}+\frac{C_{44}}{C}+\frac{2B_{4}C_{4}}{BC}=2\frac{A_{44}}{A}+\frac{A_{4}C_{4}}{AC}+\frac{A_{4}B_{4}}{AB}+\frac{16\pi G}{c^{2}}\lambda
\end{equation*}
This equation can be re-written in the following form
\begin{equation*}
    \biggl(\frac{(BC)_{4}}{BC}\biggr)_{4}+\biggl(\frac{(BC)_{4}}{BC}\biggr)^{2}=2\biggl(\frac{A_{4}}{A}\biggr)_{4}+2\frac{A^{2}_{4}}{A^{2}}+\frac{A_{4}(BC)_{4}}{ABC}
    +\frac{16\pi G}{c^{2}}\lambda
\end{equation*}
This is simplified as
\begin{equation}\label{17}
  \biggl (\frac{(BC)_{4}}{BC}-2\frac{A_{4}}{A}\biggr) _4 + \biggl (\frac{(BC)_{4}}{BC}-2\frac{A_{4}}{A}\biggr)\frac{(ABC)_{4}}{ABC}=\frac{16\pi G}{c^{2}}\lambda
\end{equation}
This equation suggests the following relation ship among A B and C
    \begin{equation}\label{18}
        A^n=BC
    \end{equation}
    where $n\neq2$, otherwise $\lambda$ would be zero.
    We can assume
    \begin{equation}\label{19}
    B=A^{n/2}D \&  C=A^{n/2}D^{-1},D=D(t)
   \end{equation}
  Further integrating equation (15), we get the first integral
\begin{equation}\label{20}
\frac{D_{4}}{D}=\frac{K}{A^{n+1}}
\end{equation}
Where K is an arbitrary constant of integration.
 Equation(13) simplifies as
 \begin{equation*}
 \rho_{4}+(n+1)\frac{A_{4}}{A}(p+\rho)-\lambda\frac{A_{4}}{A} =0
 \end{equation*}
 Using equation(7), this becomes
 \begin{equation}\label{21}
 (\rho_m)_{4}+(n+1)\frac{A_{4}}{A}(\rho_m+p) +\lambda_{4}+n\frac{A_{4}}{A}\lambda=0
 \end{equation}
 If matter and string tension co-exist without much interaction , this equation
splits into two parts.
\begin{equation}\label{22}
    (\rho_m)_{4}+(n+1)\frac{A_{4}}{A}(\rho_m+p)=0
\end{equation}
$\&$
\begin{equation}\label{23}
    \lambda_{4}+n\frac{A_{4}}{A}\lambda=0
\end{equation}
This gives following expression for matter density $\rho_m$ and string tension $\lambda$ for p=0(dust filled universe)
\begin{equation}\label{24}
\rho_m=\frac{(\rho_m)_0(A)^{n+1}_0}{A^{n+1}}
\end{equation}
and
\begin{equation}\label{25}
 \lambda=\frac{(\lambda)_0(A)^{n}_0}{A^n}
\end{equation}
The Hubble's constant in this model is
\begin{equation}\label{26}
H=u^{i}_{;i}=\frac{1}{3}
(\frac{A_{4}}{A}+\frac{B_{4}}{B}+\frac{C_{4}}{C})=\frac{(n+1)}{3}\frac{A_{4}}{A}
\end{equation}
 Equations (9)-(12) and (20) are simplified as
\begin{equation}\label{27}
\frac{n+2}{2}\frac{A_{44}}{A}+\frac{n^2}{4}\frac{A^{2}_{4}}{A^{2}}=-\frac{8\pi
G}{c^{2}}\Bigl(p-\frac{\Lambda c^{4}}{8\pi
G}+\frac{K^{2}c^{2}}{8\pi GA^{2(n+1)}}\Bigr)
\end{equation}
\begin{equation}\label{28}
\frac{A_{44}}{A}+n\frac{A^{2}_{4}}{A^{2}}=\frac{16\pi G}{c^{2}}\frac{\lambda }{n-2}
\end{equation}
\begin{equation}\label{29}
\frac{n(n+4)}{4}\frac{A^{2}_{4}}{A^{2}}=\frac{K^{2}}{A^{2(n+1)}}+\frac{8\pi G}{c^{2}}\rho+\Lambda c^{2}
\end{equation}
We see from equation(28) that ~$\lambda$=0 for ~ n=2 and other equations convert to the Einstein's field equations of LRS Bianchi type I model for perfect fluid without string\\
The average scale factor 'a' of LRS Bianchi type I model is defined as
\begin{equation}\label{30}
    a = (ABC)^{1/3}=A^\frac{n+1}{3}
\end{equation}
The spatial volume 'V' is given by
\begin{equation}\label{31}
    V = a^3 = ABC=A^{n+1}
\end{equation}
 Equations(27)-(29) can be re-written as
\begin{equation}\label{32}
    \frac{n(n+4)}{2(n+1)}\Bigl(\frac{A_{44}}{A}+\frac{n-1}{2}\frac{A^{2}_{4}}{A^{2}}\Bigr)=-\frac{8\pi G}{c^{2}}\Bigl(p-\frac{\lambda}{n+1}+\frac{K^{2}c^{2}}{8\pi GA^{2(n+1)}}-\frac{\Lambda c^{4}}{8\pi
G}\Bigr)
 \end{equation}
\begin{equation}\label{33}
 H^2=\frac{4(n+1)^2}{9n(n+4)}\frac{8\pi
G}{c^{2}}\Bigl(\rho_m+\lambda+\frac{\Lambda c^{4}}{8\pi
G}+\frac{K^{2}c^{2}}{8\pi GA^{2(n+1)}}\Bigr)
\end{equation}
We now assume that the string tension $\lambda$, cosmological constant $\Lambda$ and the
term due to anisotropy also act like energies with densities and
pressures as
$$\rho_{\lambda}=\lambda\hspace{.1in}
\rho_{\Lambda}=\frac{\Lambda c^{4}}{8\pi G}\hspace{.5in}
\rho_{\sigma}=\frac{K^{2}c^{2}}{8\pi GA^{2(n+1)}}$$
$$p_{\lambda}=-\frac{\lambda}{n+1}\hspace{.1in}
p_{\Lambda}=-\frac{\Lambda c^{4}}{8\pi G}\hspace{.5in}
p_{\sigma}=\frac{K^{2}c^{2}}{8\pi GA^{2(n+1)}}$$
\begin{equation}\label{34}
\end{equation}
It can be easily verified that energy conservation laws (22) $\&$ (23) holds
 separately for $\rho_{\Lambda}$, $\rho_\sigma$ $\&$ $\rho_{\lambda}$
 $$(\rho_{\Lambda})_{4}+3H(p_\Lambda+\rho_\Lambda)=0,~(\rho_{\sigma})_{4}+3H(p_\sigma+\rho_\sigma)=0~\&~(\rho_{\lambda})_{4}+3H(p_\lambda+\rho_\lambda)=0$$
 \begin{equation}\label{35}
  \end{equation}
  The equations of state for matter, $\lambda$,~$\sigma$ and $\Lambda$ energies are
 as follows
 $$ p_m=\omega_m\rho_m$$
$$p_{\lambda}=\omega_{\lambda}\rho_{\lambda}
\because  p_{\lambda}=-\frac{1}{n+1}\rho_{\lambda}
\therefore \omega_{\lambda}=-\frac{1}{n+1}$$
$$p_{\Lambda}=\omega_{\Lambda}\rho_{\Lambda}
 \because p_{\Lambda}+ \rho_{\Lambda}=0 \therefore \omega_{\Lambda}=-1$$
$$ \because p_{\sigma}=\rho_{\sigma}\therefore \omega_{\sigma}=1$$
\begin{equation}\label{36}
\end{equation}
where $\omega_m$=0 for matter in form of dust,
$\omega_m=\frac{1}{3}$ for matter in form of radiation. There are
certain more values of $\omega_m$ for matter in different forms
during the coarse of evolution of the universe.
Now we use the following relation between scale factor a and red
shift z
\begin{equation}\label{37}
    \frac{a_{0}}{a}=(\frac{A_{0}}{A})^{\frac{n+1}{3}}=1+z
\end{equation}
The suffix(0) is meant for the value at present time.The energy
density $\rho$ comprises of following components
\begin{equation}\label{38}
    \rho=\bigl(\rho_m+\rho_\lambda+\rho_\Lambda+\rho_\sigma \bigr)
\end{equation}
Integrations of energy equations(35) yield
$$\rho_i=(\rho_i)_0(1+z)^{3(1+\omega_i)}
\therefore \rho=\sum_i(\rho_i)_0(1+z)^{3(1+\omega_i)}$$
\begin{equation}\label{39}
\end{equation}
where suffix i corresponds to various energies densities.
We write equations (32) and (33) as
 \begin{equation}\label{40}
\frac{n(n+4)}{2(n+1)}\Bigl(\frac{A_{44}}{A}+\frac{n-1}{2}\frac{A^{2}_{4}}{A^{2}}\Bigr)=-\frac{8\pi G}{c^{2}}\Bigl(p_m+p_\lambda+p_\Lambda
+p_\sigma\Bigr)
\end{equation}
\begin{equation}\label{41}
H^{2}=\frac{4(n+1)^2}{9n(n+4)}\frac{8\pi
G}{c^{2}}\bigl(\rho_m+\rho_\lambda+\rho_{\Lambda}+\rho_{\sigma}\bigr)
\end{equation}
\section{Dust filled universe}
The present stage of the universe is full of dust for which
pressure $p_m=0$. We define critical density $\rho_{c}$ as
\begin{equation}\label{42}
 \rho_{c}=\frac{9n(n+4)c^{2}H^{2}}{32(n+1)^2\pi G}
\end{equation}
Then (41) gives
 \begin{equation}\label{43}
    \Omega_{m}+\Omega_{\lambda}+ \Omega_{\Lambda}+ \Omega_{\sigma}=1
\end{equation}
Where

$$\Omega_{m}=\frac{\rho_{m}}{\rho_{c}}=\frac{(\Omega_m)_0H^2_0(1+z)^{3}}{H^2},\Omega_{\lambda}=\frac{\rho_{\lambda}}{\rho_{c}}=\frac{(\Omega_\lambda)_0H^2_0(1+z)^{\frac{3n}{n+1}}}{H^2}$$

$$\Omega_{\Lambda}=\frac{\rho_{\Lambda}}{\rho_{c}}=\frac{(\Omega_\Lambda)_0H^2_0}{H^2},\Omega_{\sigma}=\frac{\rho_{\sigma}}{\rho_{c}}=\frac{(\Omega_\sigma)_0H^2_0(1+z)^{6}}{H^2}$$

\begin{equation}\label{44}
\end{equation}
\subsection{Expression for Hubble's Constant}
Adding all the three above and using (43), we get expression for
Hubble's constant
\begin{eqnarray*}
  H^2&=&H^{2}_0\sum_i(\Omega_i)_0(1+z)^{3(1+\omega_i)}  \\
   &=&H^{2}_0\bigl[(\Omega_m)_0(1+z)^{3}+(\Omega_\lambda)_0(1+z)^{\frac{3n}{n+1}}+(\Omega_\sigma)_0(1+z)^{6}+(\Omega_\Lambda)_0\bigr]\\
 &=&H^{2}_0\bigl[(\Omega_m)_0(\frac{A_0}{A})^{n+1}(\Omega_\lambda)_0(\frac{A_0}{A})^n+(\Omega_\sigma)_0(\frac{A_0}{A})^{2(n+1)}+(\Omega_\Lambda)_0\bigr]
\end{eqnarray*}
\begin{equation}\label{45}
\end{equation}
Now we present two graphs on the basis of above equation(see fig. 1 \& fig. 2).

These figures show that Hubble's constant increases over red shift. We can say that Hubble's constant decreases over time. We also notice from fig. 1 that in the case lambda dominated universe  ($(\Omega_m)_0$=0 or .2820 ) Hubble's constant  vary slowly  over time in comparison to matter dominated universe($\Omega_\Lambda=0 $).
\section{Luminosity Distance verses Red Shift Relation}
If x be the  spatially coordinate distance of a source with red
shift z from us, the luminosity distance which determines flux of
the source is given by
\begin{equation}\label{46}
    D_{L}=A_{0} x (1+z)
\end{equation}
Geodesic for metric (1) ensures that if in the beginning
$$\frac{dy}{ds}=0 and \frac{dz}{ds}=0$$ then
$$\frac{d^2y}{ds^2}=0 and \frac{d^2z}{ds^2}=0$$
So if a particle moves along x- direction, it continues to move
along x- direction always.If we assume that line of sight of a
vantage galaxy from us is along x-direction then path of photons
traveling through it satisfies
\begin{equation*}
ds^{2}= c^{2}dt^{2}- A^{2}dx^2=0
\end{equation*}
From this we obtain
\begin{eqnarray*}
  x&=&\int^x_{0}dx=\int^{t_{0}}_{t}\frac{dt}{A(t)}=\frac{1}{A_{0}H_{0}}\int^z_0\frac{dz}{(1+z)^{\frac{n-2}{n+1}}h(z)}  \\
  &=&\frac{1}{A_{0}H_{0}}\int^z_0\frac{dz}{(1+z)^{\frac{n-2}{n+1}}\sqrt{\bigl[(\Omega_m)_0(1+z)^3+(\Omega_\lambda)_0(1+z)^{\frac{3n}{n+1}}+(\Omega_\sigma)_0(1+z)^6+(\Omega_\Lambda)_0\bigr]}}
\end{eqnarray*}
\begin{equation}\label{47}
\end{equation}
Where we have used $ dt=dz/\dot{z}$ and from eqns (26) and (37)
$$\dot{z}=-H(1+z)~\&~ h(z)=\frac{H}{H_0}$$.\\
So the luminosity distance is given by
\begin{equation}\label{48}
D_{L}=\frac{c(1+z)}{H_{0}}\int^z_0\frac{dz}{(1+z)^{\frac{n-2}{n+1}}\sqrt{\bigl[(\Omega_m)_0(1+z)^3+(\Omega_\lambda)_0(1+z)^{\frac{3n}{n+1}}+(\Omega_\sigma)_0(1+z)^6+(\Omega_\Lambda)_0\bigr]}}
\end{equation}
\section{Apparent Magnitude verses Red Shift relation for Type Ia supernova's(SN Ia):} The Absolute and Apparent magnitude of a source (M and m) are
related to the red shift of the source by following relation
\begin{equation}\label{49}
    m-M = 5log_{10}\bigl(\frac{D_L}{Mpc}\bigr)+25
\end{equation}
The Type Ia supernova (SN Ia) can be observed when white dwarf
stars exceed the mass of the Chandrasekhar limit and explode. The
belief is that SN Ia are formed in the same way irrespective of
where they are in the universe, which means that they have a
common absolute magnitude M independent of the red shift z. Thus
they can be treated as an ideal standard candle. We can measure
the apparent magnitude m and the red shift z observationally,
which of course depends upon the objects we observe.\\ To get
absolute Magnitude M of a supernova, we consider supernova at very
small red shift. Let us consider a supernova 1992P at low-red
shift z = 0.026 with m = 16.08.For low red shift supernova,we have
the following relation
\begin{equation}\label{50}
D_L=\frac{cz}{H_0}
\end{equation}
Putting values z = 0.026, m = 16.08 and $D_L$ from (50),\\
Equation (49) gives M for all SN Ia as follows.
\begin{equation}\label{51}
    M = 5log_{10}\bigl(\frac{H_0}{.026 c}\bigr)-8.92
\end{equation}
From equations (49) and (51)
\begin{equation}\label{52}
 log_{10}(H_{0}D_{L})= (m - 16.08)/5 + log_{10}(.026c Mpc)
\end{equation}
The equation(52) gives observed value of Luminosity distance in term of observed value of apparent magnitude whereas the equation (48) gives the theoretically  value of luminosity distance in terms of red shift on the basis of our model.
Next, from equations (52) and (48) we get the following expression for (m,z)
relations of Supernova's SN Ia.
\[
 m=16.08+\\5log_{10}\Bigl(\frac{1+z}{.026}\Bigr)\times
\]
\begin{equation}\label{53}
\int^z_0\frac{dz}{(1+z)^{\frac{n-2}{n+1}}\sqrt{\bigl[(\Omega_m)_0(1+z)^3+(\Omega_\lambda)_0(1+z)^{\frac{3n}{n+1}}+(\Omega_\sigma)_0(1+z)^6+
(\Omega_\Lambda)_0\bigr]}}
\end{equation}
\section{Estimation of Present values of Energy parameters $(\Omega_m)_0,(\Omega_\Lambda)_0,(\Omega_\lambda)_0 ~\&~ (\Omega_\sigma)_0$ }
String tension play important role in the study of early evolution of the universe.
We also notice that after publication of WMAP data, today there is considerable evidence in support of
anisotropic model of universe. In the past, there might had been certain anisotropies in the universe .
Our model validates that both string tension as well as anisotropy decreases over time. Considering it, we  give very low present values to
$(\Omega_\sigma)_0$ and $(\Omega_\lambda)_0$. In our earlier work  we developed models with n = 2 which make string tension $\lambda$=0. So we take
\begin{equation}\label{54}
    n=2.1,(\Omega_\sigma)_0=.0002 \& (\Omega_\lambda)_0=.0001
\end{equation}
 in our present model.
 Now we present two Tables(See Appendix-1 and 2) which
contain $287$ high red shift ($ 0.3 \leq z \leq 1.4$ ) SN Ia
supernova data of observed apparent magnitude and luminosity distances along with their
possible error from Union $2.1$ compilation. These Tables also
contains the corresponding theoretical values of apparent
magnitudes and luminosity distances  for $\Omega_{m}$ = $0.2820$, $0$ and $1$ respectively
which are obtained as per from Equation(48) and (52). As stated in
the introduction, our purpose is to obtain the  results close to WMAP on
the basis of union 2.1 compilations for our model, so we have obtained the
various sets of theoretical data's of apparent magnitudes  and luminosity distances
corresponding to different values of $\Omega_{m}$ in between $0$
to $1$. In order to see that out of the these sets of theoretical data's of apparent magnitudes
and luminosity distances which one is close to the observational values, we calculate
$\chi^{2}$ using following formula due to Amanullah et al.(2010).
The tables[1] and [2] describe the various values of  $\chi^{2}$
against values of $\Omega_{m}$ ranging in between $0$ to $1$. $\vphantom{}$
\begin{center}
{\large{$\chi_{SN}^{2}=A-\frac{B^{2}}{C}+log_{10}(\frac{C}{2\pi})$}}\par
\end{center}{\large \par}
{\large{$ $}}{\large \par}
Where
\begin{center}
{\large{$A=\overset{60}{\underset{i=1}{\sum}}\frac{\left[\left(m\right)_{ob}-\left(m\right)_{th}\right]^{2}}
{\sigma_{i}^{2}}$}}\par
\end{center}
{\large \par}
\begin{center}
{\large{$B=\overset{60}{\underset{i=1}{\sum}}\frac{\left[\left((m\right)_{ob}-\left((m\right)_{th}\right]}
{\sigma_{i}^{2}}$}}\par
\end{center}
{\large \par}
And
\begin{center}
{\large{$C=\overset{60}{\underset{i=1}{\sum}}\frac{1}{\sigma_{i}^{2}}$}}\par
\end{center}
{\large \par}
\begin{equation}\label{55}
\end{equation}
\begin{center}
\begin{tabular}{ | c | c | c | c | c | } \hline
$(\Omega_{m)_0}$ &    $ \chi_{SN}^{2}$ & $\chi_{SN}^{2}/287$\tabularnewline
\hline
0	&	5462.2	&	19.03205575	 \tabularnewline
\hline
0.1	&	4972.8	&	17.32682927	 \tabularnewline
\hline
0.2	&	4822.8	&	16.80418118	 \tabularnewline
\hline
0.21	&	4816.5	&	16.78222997	 \tabularnewline
\hline
0.26	&	4800.2	&	16.72543554	 \tabularnewline
\hline
0.27	&	4799.4	&	16.72264808	 \tabularnewline
\hline
0.28	&	4799.3	&	16.72229965	 \tabularnewline
\hline
0.28	&	4799.3	&	16.72229965	 \tabularnewline
\hline
0.281	&	4799.3	&	16.72229965	 \tabularnewline
\hline
0.282	&	4799.3	&	16.72229965	 \tabularnewline
\hline
0.283	&	4799.4	&	16.72264808	 \tabularnewline
\hline
0.29	&	4799.8	&	16.72404181	 \tabularnewline
\hline
0.3	&	4800.9	&	16.72787456	 \tabularnewline
\hline
0.9997	&	5312.2	&	18.50940767	 \tabularnewline
\hline
\end{tabular}
\end{center}
\begin{center}
\textbf{{Table 1 :~}{\small{$\chi^2$ table for best fitting theoretical and observed values of apparent magnitudes`m'}}}
\end{center}
\newpage
\begin{center}
\begin{tabular}{ | c | c | c | c | c | } \hline
$(\Omega_{m)_0}$ &    $ \chi_{SN}^{2}$ & $\chi_{SN}^{2}/287$\tabularnewline
\hline
0	&	223.1983	&	0.777694425	 \tabularnewline
\hline
0.1	&	203.6246	&	0.70949338	 \tabularnewline
\hline
0.2	&	197.6239	&	0.688585017	 \tabularnewline
\hline
0.21	&	197.3738	&	0.687713589	 \tabularnewline
\hline
0.26	&	196.7195	&	0.685433798	 \tabularnewline
\hline
0.27	&	196.6878	&	0.685323345	 \tabularnewline
\hline
0.28	&	196.6836	&	0.685308711	 \tabularnewline
\hline
0.281	&	196.6836	&	0.685308711	 \tabularnewline
\hline
0.282	&	196.6836	&	0.685308711	 \tabularnewline
\hline
0.283	&	196.6874	&	0.685321951	 \tabularnewline
\hline
0.29	&	196.7049	&	0.685382927	 \tabularnewline
\hline
0.3	&	196.7499	&	0.685539721	 \tabularnewline
\hline
0.9997	&	217.2009	&	0.756797561	 \tabularnewline
\hline
\end{tabular}
\end{center}
\begin{center}
  \textbf{{Table 2 :~}{\small{$\chi^2$ table for best fitting theoretical and observed values of luminosity distance$D_{L}$}}}
\end{center}
The above tables depict the fact that the best fit values of $\Omega_m$ and therefore $\Omega_\Lambda$
are as follows
\begin{equation}\label{56}
 \Omega_m= 0.2820~~ \&~~ \Omega_\Lambda= 0.7177
\end{equation}
\subsection{Deceleration Parameter}
The deceleration parameter is given by
\begin{equation*}
q=-\frac{aa_{44}}{a_{4}^2}= -\frac{n-2}{n+1}-\frac{3}{n+1}\frac{AA_{44}}{A_{4}^{2}}
\end{equation*}
 From equations (39)-(53)
   \begin{eqnarray*}
    \frac{AA_{44}}{A^2_4}&=& -\frac{n+1}{2}\sum_{i}\omega_{i}\Omega_{i}-\frac{n-1}{2} \\
   &=& -\frac{n+1}{2}(-\Omega_{\Lambda}-\frac{1}{n+1}\Omega_{\lambda}+\Omega_{\sigma})-\frac{n-1}{2}\\
   \therefore q &=& \frac{3}{2}\frac{(-(\Omega_\Lambda)_0-\frac{1}{n+1}(\Omega_\lambda)_0(1+z)^{\frac{3n}{n+1}}+(\Omega_\sigma)_0(1+z)^6)}
   {\bigl[(\Omega_m)_0(1+z)^{3}+(\Omega_\lambda)_0(1+z)^{\frac{3n}{n+1}}+(\Omega_\sigma)_0(1+z)^{6}+(\Omega_\Lambda)_0\bigr]}+\frac{1}{2}
   \end{eqnarray*}
   \begin{equation}\label{57}
    \end{equation}
 $$ At~~z=0.7170    ,~~ q= -0.0004 ~~\&~~at~~ z= 0.7180,~~ q=.0002$$
  So the universe entered the accelerating phase at
$z\thicksim0.7175\backsimeq t\thicksim 0.4616 H^{-1}_0\thicksim 6.27\times10^9 yrs$ in the past
before as from now.\\
The present value of q is
 $$q_0=-0.5793$$
This equation clearly show that without presence of $\Lambda$
   term in the Einstein's field equation(\ref{8}), one can't imagine of accelerating
   universe.This equation also expresses the fact that anisotropy
   raises the lower limit value of $(\Omega_\Lambda)_0$ required for acceleration at present. This
   may be seen in the following way.\\
   For isotropic model(FRW) acceleration requires\\
   \begin{equation}\label{58}
  ( \Omega_{\Lambda})_0\geq.33
\end{equation}
 For anisotropic model with out string
\begin{equation}\label{59}
(\Omega_{\Lambda})_0\geq.33+(\Omega_{\sigma})_0=.3302
\end{equation}
 For  anisotropic model with  string
 \begin{equation}\label{60}
(\Omega_{\Lambda})_0\geq.3559+(\Omega_{\sigma})_0 -.0123(\Omega_{\lambda})_0=.3360
\end{equation}
\subsection{Age of the Universe}
The present age of the universe is obtained as follows
\begin{eqnarray*}
   t_{0}&=&\int^{t_0}_{0}dt=\int^{\infty}_{0}\frac{dz}{H(1+z)} \\
   &=& \int^{\infty}_0\frac{dz}{H_0 (1+z)\sqrt{\bigl[(\Omega_m)_0(1+z)^3+(\Omega_\lambda)_0(1+z)^{\frac{3n}{n+1}}+(\Omega_\sigma)_0(1+z)^6+(\Omega_\Lambda)_0\bigr]}} \\
   &=&\int^{\infty}_0\frac{3dA}{(n+1)H_0A\sqrt{\bigl[(\Omega_m)_0(\frac{A_0}{A})^{(n+1)}+(\Omega_\lambda)_0(\frac{A_0}{A})^{n}+(\Omega_\sigma)_0(\frac{A_0}{A})^{2(n+1)}+(\Omega_\Lambda)_0\bigr]}} \\
\end{eqnarray*}
\begin{equation}\label{61}
\end {equation}
Where we have used $ dt=dz/\dot{z}$ and  $\dot{z}=-H(1+z)$\\
We see that at large red-shift, $H_0 t$ becomes constant and it tends to the value $0.9505$. So
the Present age of the universe as per our model is\\
\begin{equation}\label{62}
    t_0 =  0.9505 H^{-1}_0= 13.2799\times10^9 years
\end{equation}
$ $
\subsection{Densities of the various Energies in the universe :}
$ $
The energy density  is given by (39)
$$\rho_i=(\Omega_i)_0 (\rho_c)_0 (1+z)^{3(1+\omega_i)}$$
$$\therefore \rho=\sum_i(\Omega_i) (\rho_c)_0 (1+z)^{3(1+\omega_i)}$$
\begin{equation}\label{63}
\end{equation}
where i stands for various types of energies such as matter energy, dark energy,energies due to anisotropy of the universe and string tension.
Taking,n=2.1,$\Omega_m=0.2820,\Omega_{\Lambda}=0.7177,\Omega_{\sigma}=0.0002 \& \Omega_\lambda=0.0001$ , $H_{0}=72km/sec./Mpc$,$h_0\backsimeq.72$
$G=6.6720e-008 cm^{3}/sec^{2}/gm$
\begin{equation}\label{64}
 (\rho_c)_0   =\frac{9n(n+4)c^{2}H_{0}^{2}}{32(n+1)^2\pi G}=1.8795h^2_0\times10^{-29}gm/cm^{3}
\end{equation}
The dust energy and its current value  for flat universe is given as
\begin{equation}\label{65}
(\rho_{m})=(\Omega_m)_0 (\rho_c)_0(1+z)^3,~\&~(\rho_{m})_{0}=.5262  h^2_0\times10^{-29}gm/cm^{3}
\end{equation}
  dark energy $(\rho_{\Lambda})$ and its current value is
given as
\begin{equation}\label{66}
(\rho_{\Lambda})=(\rho_{\Lambda})_{0}=1.35268 h^2_0\times10^{-29}gm/cm^{3}
\end{equation}
The string tension $\lambda$ and its current value is given as
\begin{equation}\label{67}
\lambda=(\Omega_\lambda)_0 (\rho_c)_0(1+z)^{2.0323}~\&~(\lambda)_0=(\rho_{\lambda})_{0}=0.00018 h^2_0\times10^{-29}gm/cm^{3}
\end{equation}
And finally anisotropy $\rho_{\sigma}$ and its current value is given as
\begin{equation}\label{70}
(\rho_{\sigma})=(\Omega_{\sigma})_0 (\rho_c)_0 (1+z)^6~\&~(\rho_{\sigma})_{0}=0.00036h^2_0\times10^{-29}gm/cm^{3}
\end{equation}
\begin{figure*}[thbp]
\begin{tabular}{rl}
\includegraphics[width=7.5cm]{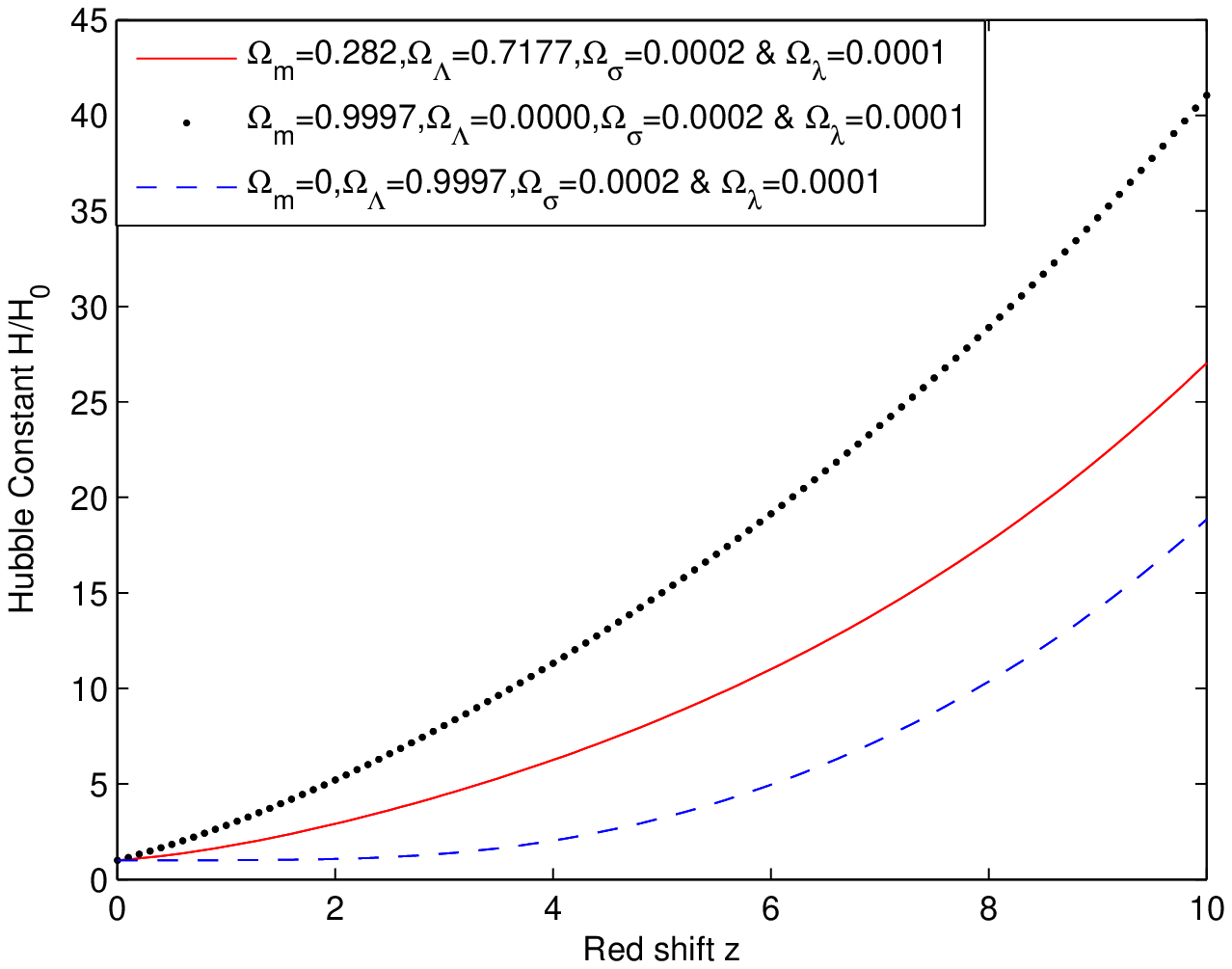}&
\includegraphics[width=7.5cm]{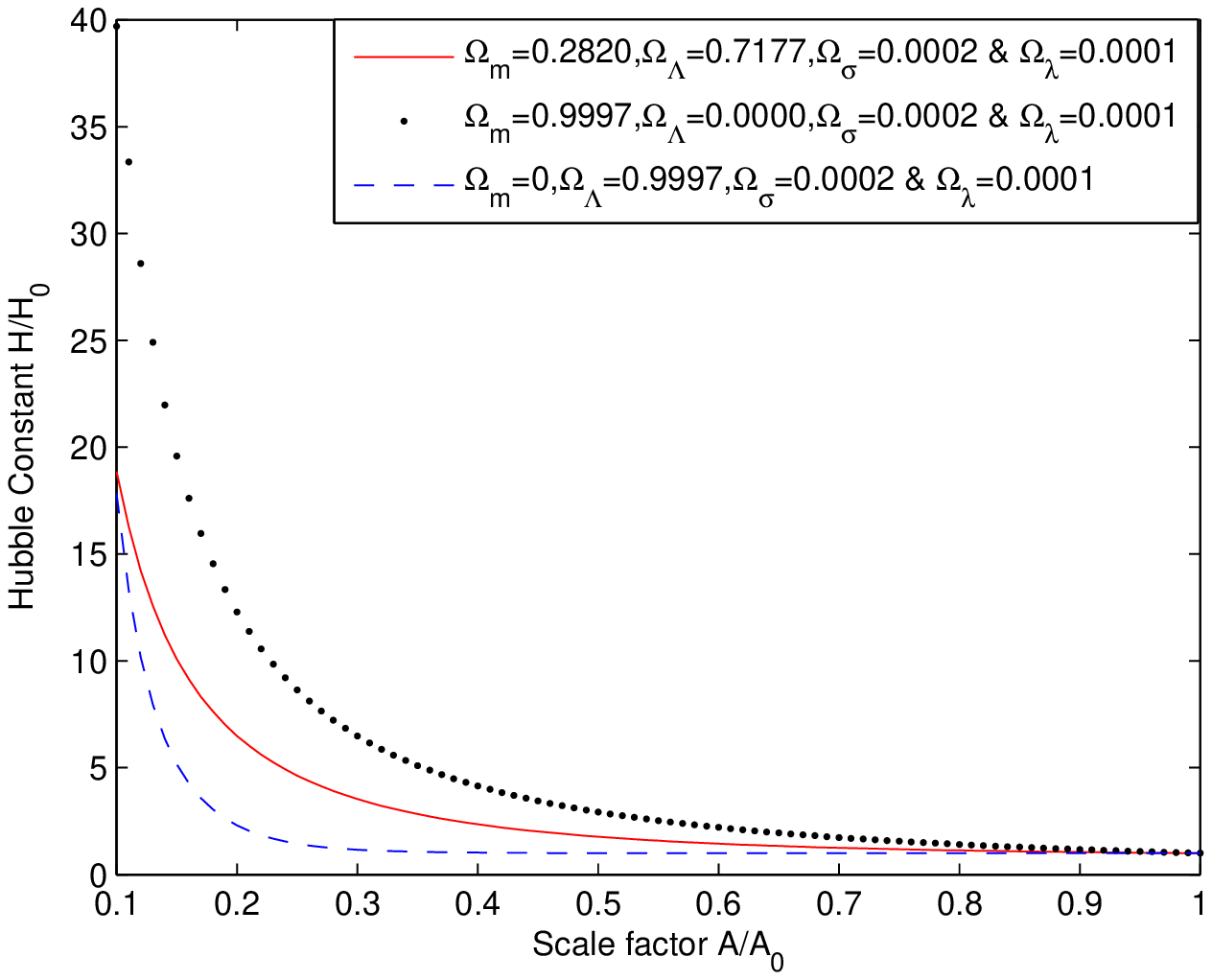}\\
\includegraphics[width=7.5cm]{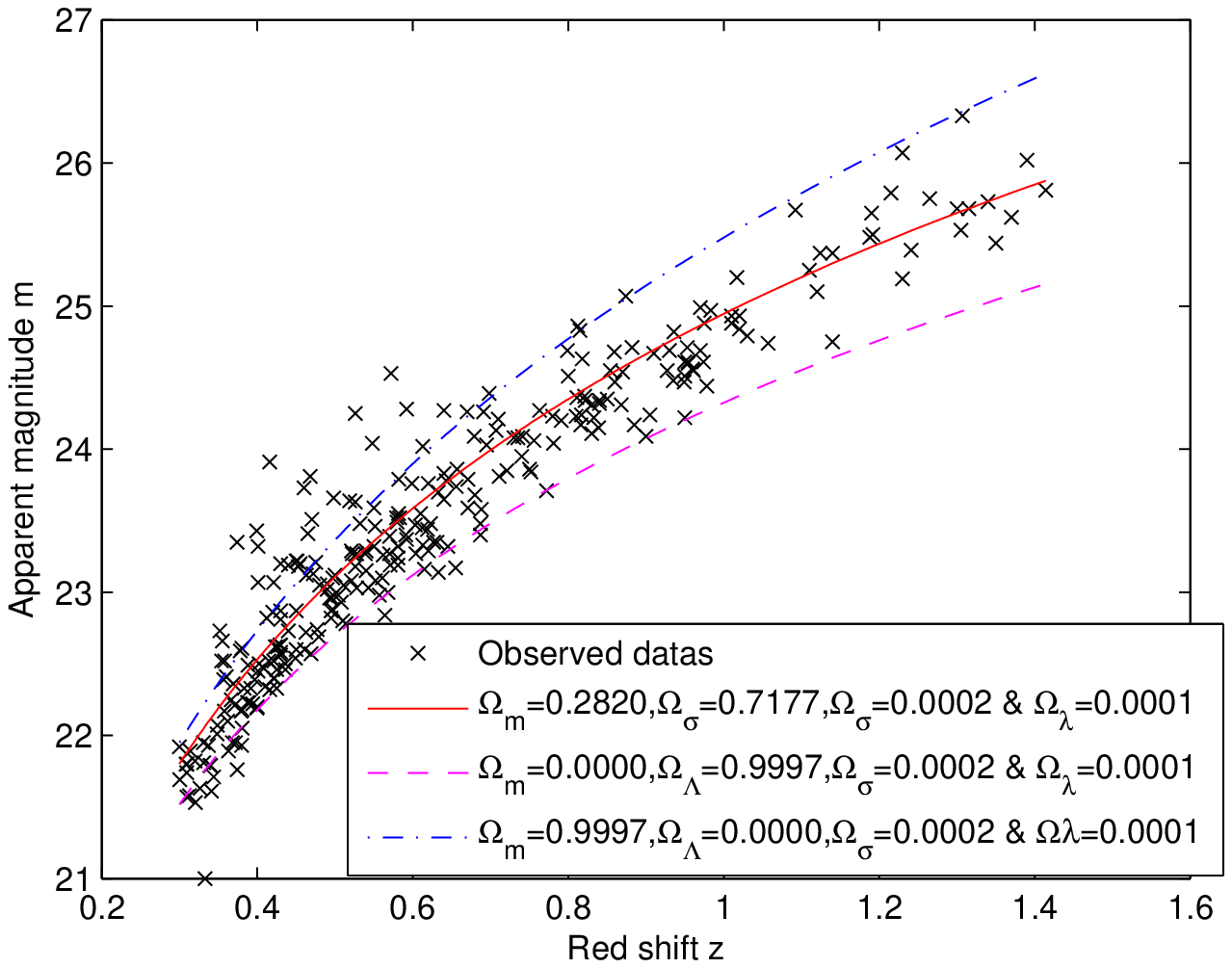}&
\includegraphics[width=7.5cm]{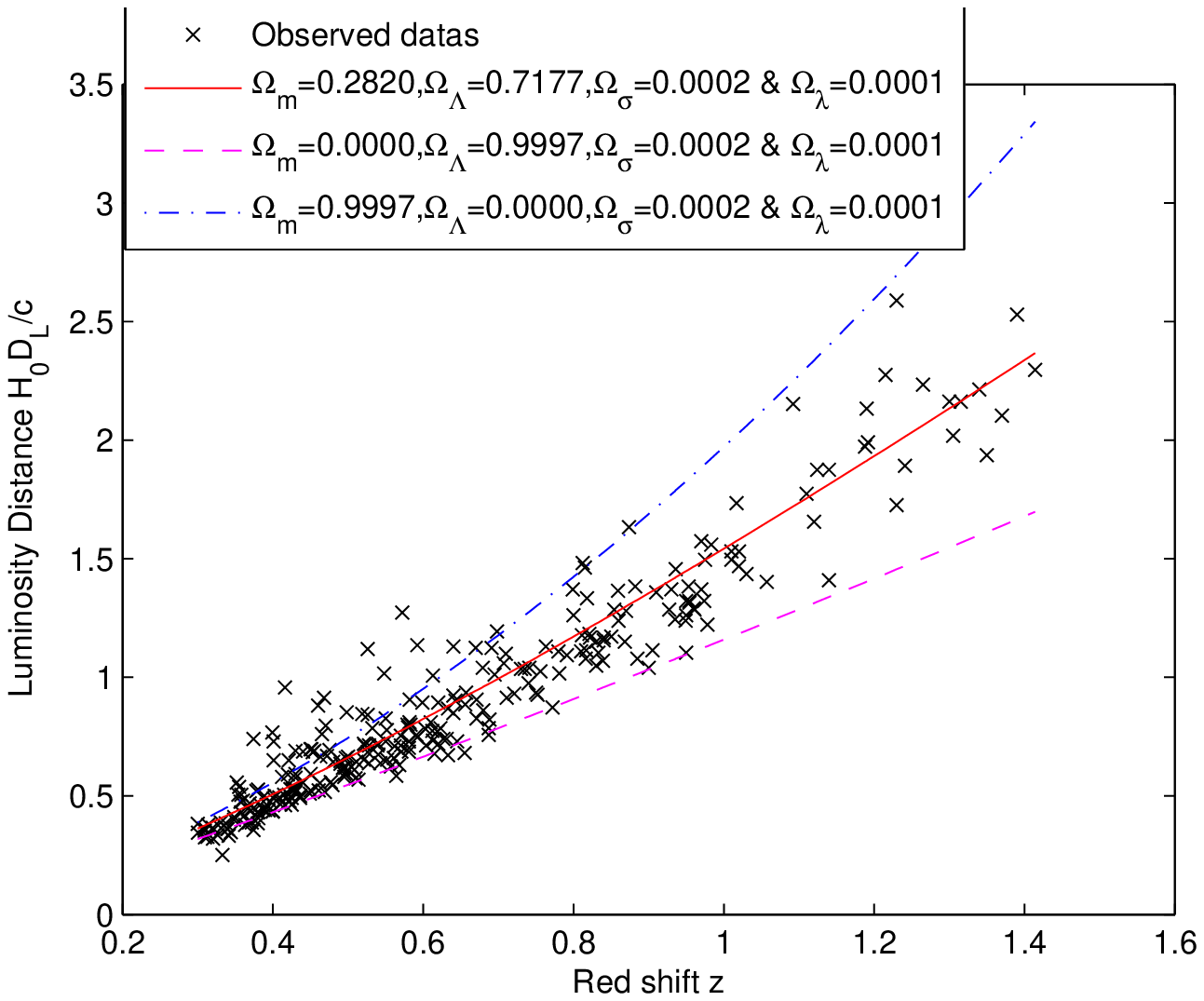}\\
\includegraphics[width=7.5cm]{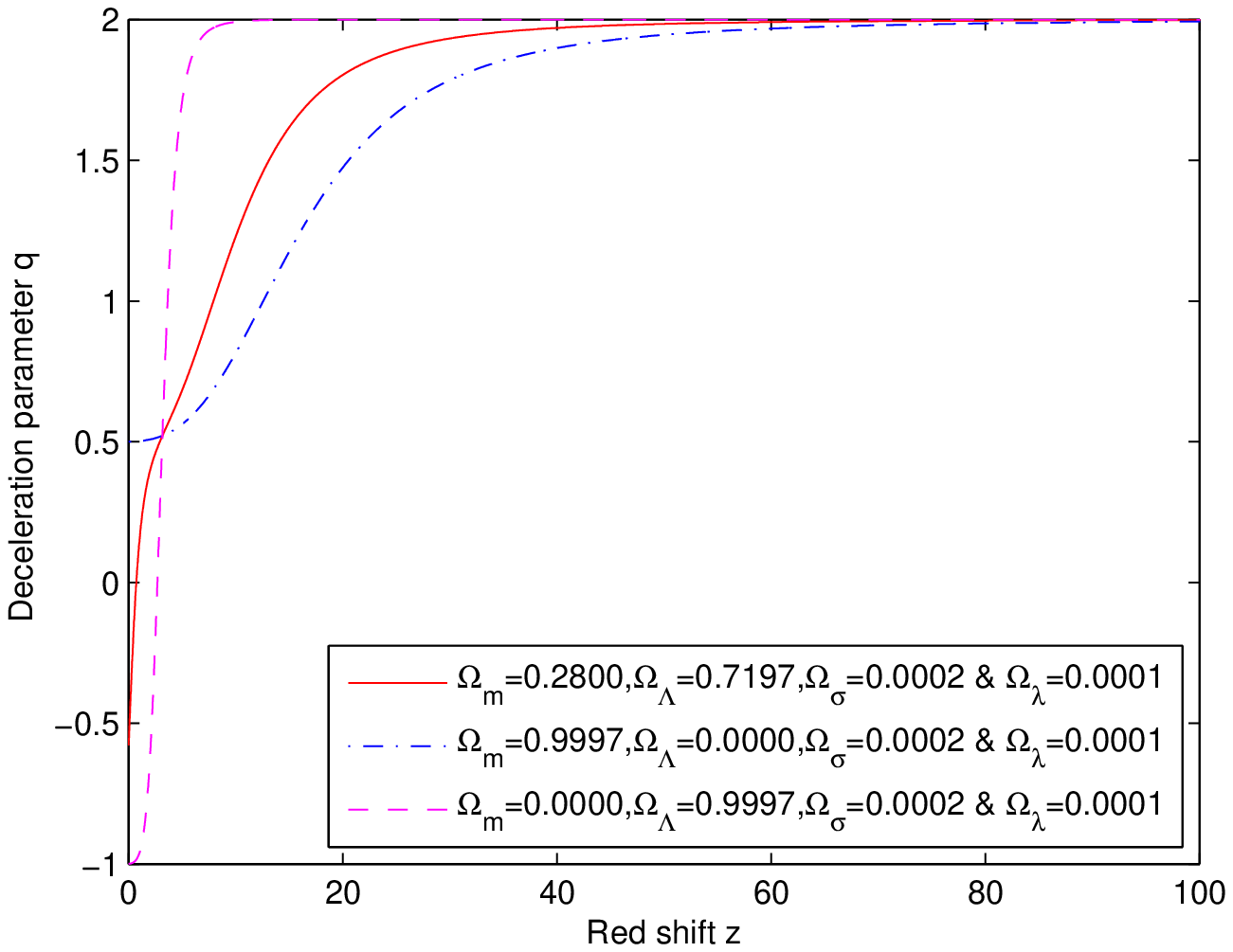}&
\includegraphics[width=7.5cm]{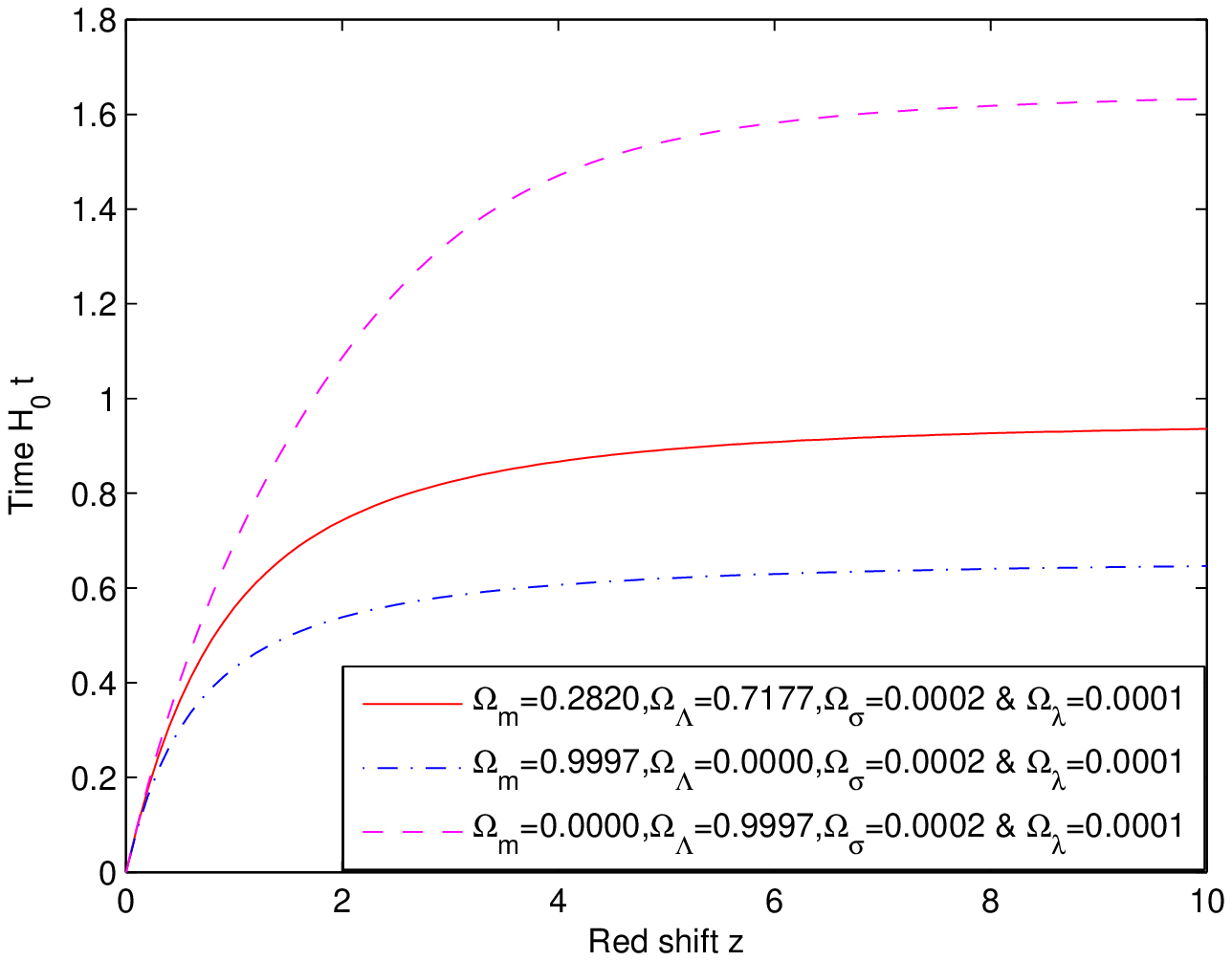}\\
\includegraphics[width=7.5cm]{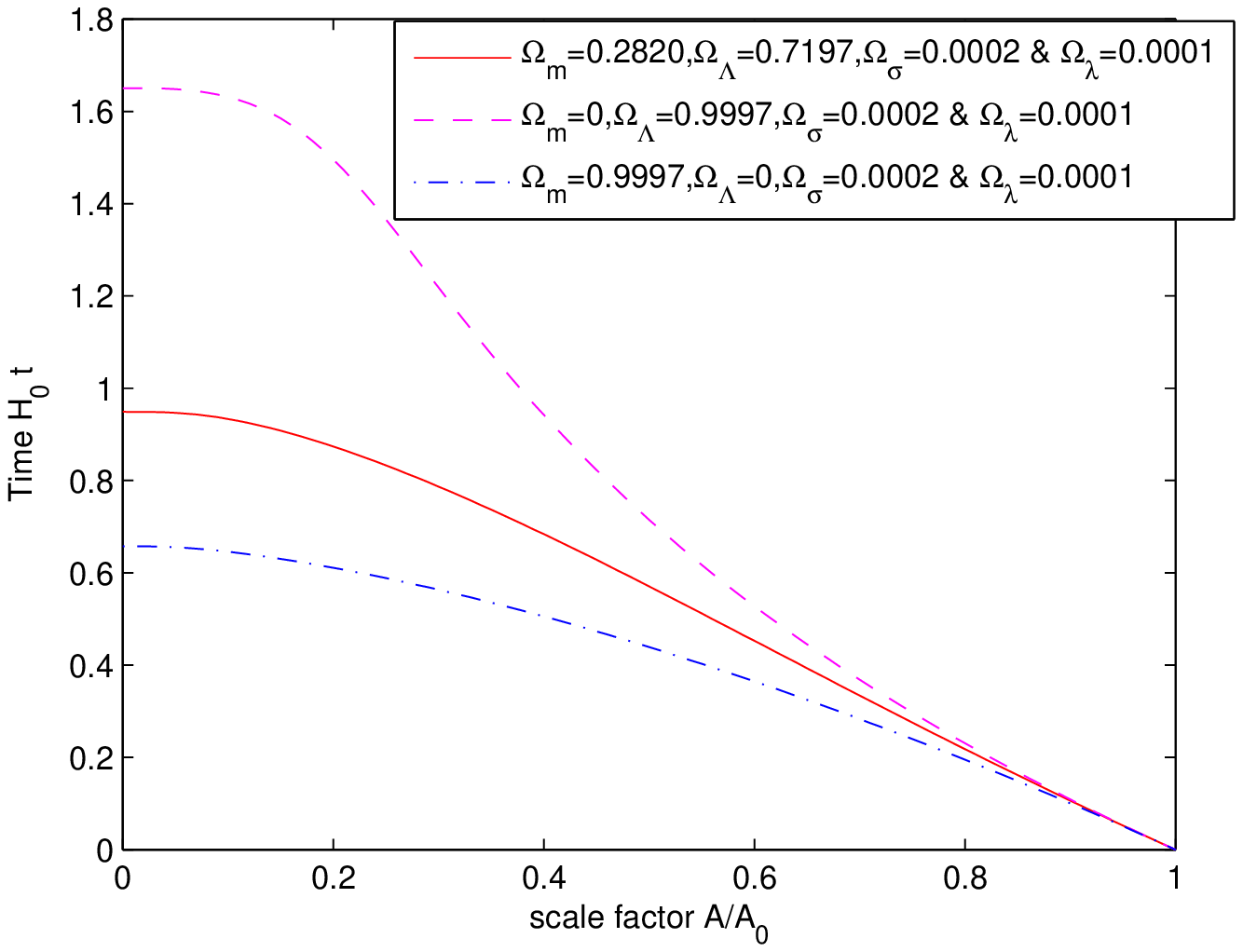}&
\includegraphics[width=7.5cm]{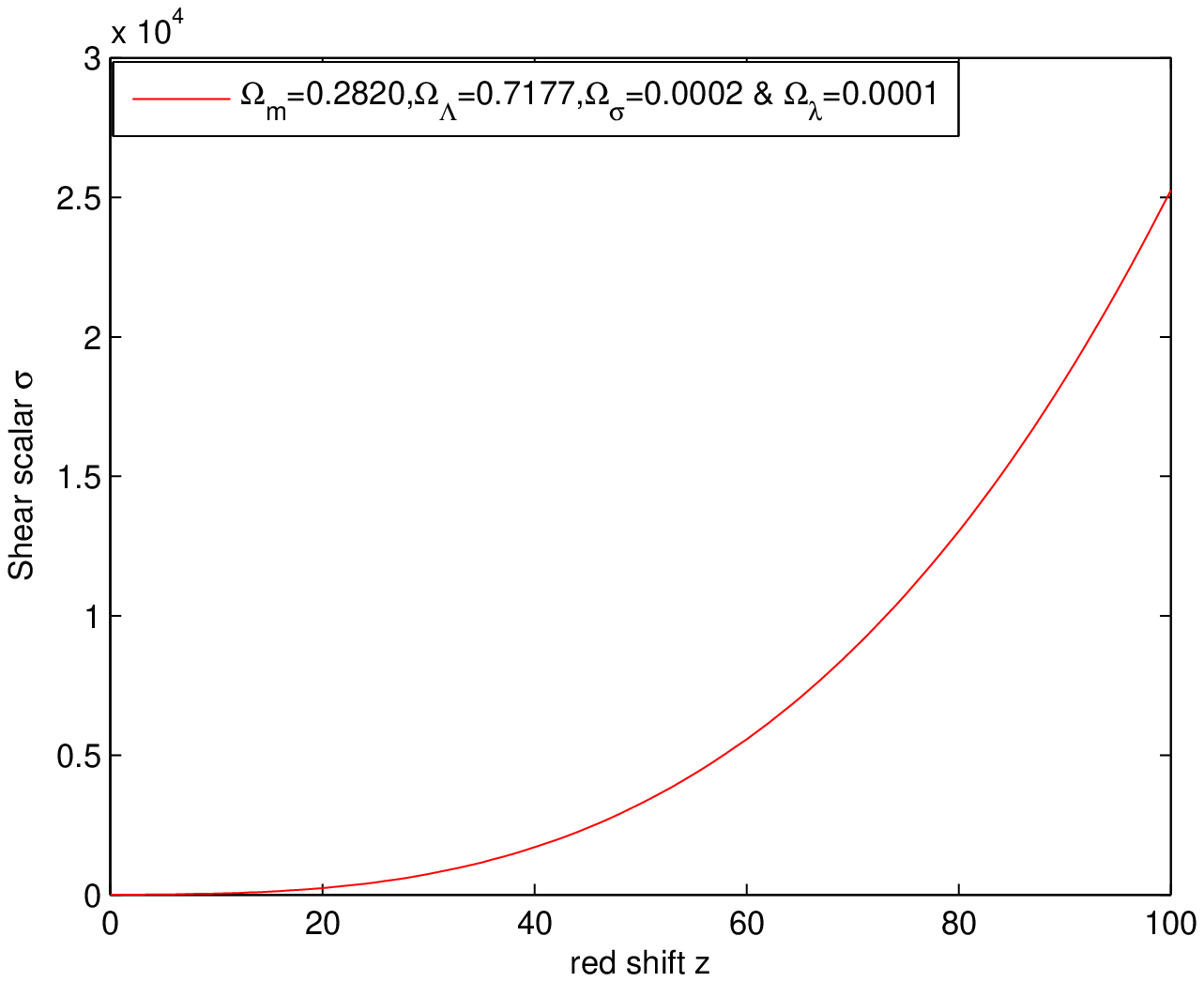}
\end{tabular}
\caption{Variation of $\frac{H}{H_{0}}$, $m$, $\frac{H_{0}D_{L}}{c}$, $q$, $H_{0}t$ and $\sigma$ with redshift.}
\end{figure*}
\subsection{Shear Scalar($\sigma$), Relative Anisotropy($\sigma^2/\rho_m$) \&  Mean Anisotropy Parameter ($A_{m}$)}
 The shear scalar($\sigma$), relative anisotropy($\sigma^2/\rho_m$) \&  the mean anisotropy parameter
($A_{m}$)are defined as and are given by\\
 \begin{eqnarray*}
   \sigma^{2}&=&\frac{1}{2}(\overset{3}{\underset{i=1}{\sum}}H_{i}^{2}-3\theta^{2})\\
   &=&\frac{(n-2)^2}{12}\frac{A^{2}_{4}}{A^2}+\frac{D^{2}_{4}}{D^2}\\
   \sigma&=&H_0\sqrt{\frac{3(n-2)^2}{4(n+1)^2}\bigl[(\Omega_m)_0(1+z)^3+(\Omega_\lambda)_0(1+z)^{\frac{3n}{n+1}}+(\Omega_\Lambda)_0\bigr]+3(\Omega_\sigma)_0(1+z)^6}
   \end{eqnarray*}
 \begin{equation}\label{72}
 \end{equation}
 \begin{equation}\label{73}
    \frac{\sigma^{2}}{\rho_m}=\frac{H^2_0\big(\frac{3(n-2)^2}{4(n+1)^2}\bigl[(\Omega_m)_0(1+z)^3+(\Omega_\lambda)_0(1+z)^{\frac{3n}{n+1}}+(\Omega_\Lambda)_0\bigr]+3(\Omega_\sigma)_0(1+z)^6\bigr)}{(\Omega_m)_0 (\rho_c)_0(1+z)^3}
 \end{equation}
\begin{eqnarray*}
A_m&=&\frac{1}{3}\overset{3}{\underset{i=1}{\sum}}(\frac{H_{i}-H}{H})^2\\
&=&\frac{(n-2)^2}{2(n+1)^2}+\frac{6A^2D_4^2}{(n+1)^2D^2 A^2_4}\\
&=&\frac{(n-2)^2}{2(n+1)^2)}+\frac{3n(n+4)}{2(n+1)^2}\frac{(\Omega_\sigma)_0(1+z)^6}{\bigl[(\Omega_m)_0(1+z)^{3}+(\Omega_\lambda)_0(1+z)^{\frac{3n}{n+1}}+(\Omega_\sigma)_0(1+z)^{6}+(\Omega_\Lambda)_0\bigr]}
\end{eqnarray*}
\begin{equation}\label{74}
\end{equation}
where $ H_{i}(i=1,2,3)$ represent the directional Hubble parameters in the direction of x,y and z , respectively.\\
The present value of shear scalar is given as\\
$$ (A_m)_0= 0.38107518$$
The present value of mean anisotropy parameter $A_m$ is given as\\
$$ (\sigma)_0= 1.1588H_0=11.84\times10^{-11}yrs^{-1}$$
\vspace{10mm}
\section{Conclusion}
We summarize our work by presenting the
following table which displays the values of cosmological
parameters at present epoch.
\begin{center}
\begin{tabular}{|c|c|c|c|c|}
  \hline
  Cosmological Parameters & Values at Present \\
  \hline
  $(\Omega_\Lambda)_0$ & .7177 \\
   $(\Omega_m)_0$ & .2820\\
   $(\Omega_\lambda)_0$ & .0001 \\
   $(\Omega_\sigma)_0$ & .0002 \\
  $(q)_0$& -0.5793\\
  $\small{Matter~ energy~ density}= (\rho_{m})_{0}$ &$0.5262 h^2_0\times10^{-29}gm/cm^{3}$  \\
  $\small{Dark~ energy~ density}=(\rho_{\Lambda})_{0}$& $1.35268h^2_0\times10^{-29}gm/cm^{3}$\\
  $\small{String~ tension}=(\lambda)_0$&$0.00018 h^2_0\times10^{-29}gm/cm^{3}$\\
  $(\sigma)_0$ & $1.1588H_0=11.84\times10^{-11}yrs^{-1}$\\
  $(A_m)_0$& $0.038107518$\\
\hline
\end{tabular}
\end{center}

Also we observe that acceleration begun before $6.27 \times 10^{9}$ years and the present age
of universe is 13.2799 Gyears.
\subsection*{Acknowledgements}
This work is  supported by the CGCOST Minor Research Project
789/CGCOST/MRP/14.

\end{document}